\documentstyle[prd,aps]{revtex}
\tighten
\begin{document}
\draft
\title{     Periodic Solutions of the Einstein Equations
            for Binary Systems}
\author{    Steven Detweiler\footnote{Email: det@phys.ufl.edu}}

\address{   Department of Physics,
            University of Florida,
            Gainesville, FL 32605
        }
\maketitle
\begin{abstract}

Solutions of the Einstein equations which are periodic and have
standing gravitational waves, in the weak-field zone, are
valuable approximations to more physically realistic solutions
with outgoing waves. A variational principle for the periodic
solutions is found which has the power to provide, for binary
systems with weak gravitational radiation, an accurate estimate
of the relationship between the mass and angular momentum of the
system, the masses and angular momenta of the components, the
rotational frequency of the frame of reference in which the
system is periodic, the frequency of the periodicity of the
system, and the amplitude and phase of each multipole component
of gravitational radiation. Examination of the boundary terms of
the variational principle leads to definitions of the effective
mass and effective angular momentum of a periodic geometry which
capture the concepts of mass and angular momentum of the source
alone with no contribution from the gravitational radiation.
These effective quantities are surface integrals in the
weak-field zone which are independent of the surface over which
they are evaluated, through second order in the deviations of the
metric from flat space. The variational principle provides a
powerful method to examine the evolution of, say, a binary black
hole system from the time when the holes are far apart, through
the stage of slow evolution caused by gravitational radiation
reaction, up until the moment when the radiation reaction
timescale is comparable to the dynamical timescale.
\end{abstract}


\section{Introduction}

The difficulties associated with the study
of solutions of the Einstein equations are well known.  However,
a number of specialized techniques have been developed which
allow analyses in rather narrow circumstances.  For example, the
post-Newtonian approximation yields relativistic corrections for
systems which are restricted to slow speeds and weak
gravitational fields. Or, symmetric geometries with two or three
Killing vectors allow for study of Kerr or Schwarzschild black
holes and both rotating and non-rotating neutron star models.  A
particularly fruitful technique involves the perturbation
analysis of analytically known solutions---perturbations of flat
space comprise linearized gravity\cite{ThorneRMP}; the
perturbations of spherically symmetric geometries are used to
study the emission of radiation from test particles orbiting a
black hole and the quasi-normal oscillations of black
holes\cite{MTBH} and neutron stars\cite{ThorneII}.

This paper presents a new restriction of the Einstein equations
which is sufficiently limiting that analysis can proceed and yet
sufficiently general to encompass a wide variety of interesting
applications.  Interest focuses on those solutions of the
Einstein equations which allow a coordinate system in which the
geometry is periodic in time inside a bounded region of
space-time.  Such a geometry might have strong fields,
gravitational waves and high speeds; and it might involve black
holes or neutron stars.

In quantum mechanics the physically realistic solutions of the
Schr\"odinger equation with, say, an incoming wave
packet being scattered into an outgoing wave packet are
constructed from a linear combination of the periodic
solutions which contain standing waves.  Thus, the scattering
problem is reduced to the analysis of periodic solutions.

In general relativity the nonlinearity of the Einstein equations
intrudes on using this same idea of constructing physically
interesting solutions from the periodic ones.  But with care and
some limitations, the basic process still provides an
avenue toward otherwise unapproachable physical systems.

Periodic solutions of the Einstein equations with radiation are
not asymptotically flat\cite{gibbons84}---this is not surprising:
in the weak-field zone, where the linearized Einstein equations
give an approximate description of the gravitational field, the
mass density of standing waves falls off as $r^{-2}$. Thus, the
contribution of radiation to the mass content inside a radius $r$
grows linearly with $r$; and the linearized Einstein equations
cannot give an accurate description of the gravitational field
out to either spatial or null infinity. But this is {\em not} a
critical limitation to periodic geometries.

In this paper much attention is focused on the ``weak-field
zone''---a region of space-time, surrounding a fully relativistic
source, throughout which the linearized Einstein equations give
an accurate description of the gravitational field.  A further
requirement which we impose upon our use of the weak-field zone
is that its total energy content in gravitational waves be much
less than the mass of the source. Thus the weak-field zone is
restricted to a region where the nonlinear effects of the
radiation are small. For modeling realistic binary systems the
boundary of our region of interest is always within the
weak-field zone.  Thus, {\em for periodic geometries we consider
only a bounded region of space, and describe and impose boundary
conditions well away from spatial or null infinity}---which might
not even exist for some continuations past the boundary of a
periodic geometry!

For a periodic example, consider the binary pulsar---the neutron
stars themselves are relativistic, collapsed objects; and the
system has an orbital period of a bit less than eight hours and a
total mass of about \(3\text{M}_{\odot}\).  A region bounded at a
distance of one light month contains approximately 200
wavelengths of quadrupole radiation from the source; \(m/r\sim
10^{-11}\); the approximate energy content in the gravitational
waves out to this distance is $10^{-22}\text{M}_{\odot}$; and the
approximate binding energy of the system is $Gm^2/R\sim
7\times10^{-7}\text{M}_{\odot}$. The true, physical system has
outgoing radiation near the boundary, and the orbit decays as a
consequence. But a similar bounded, periodic, exact solution to
the Einstein equations could be constructed from this physically
realistic one by sending gravitational radiation inward from the
boundary with amplitude and phase chosen to keep the system from
evolving.  The nature of the periodic geometry outside the
boundary need not be considered.  However, if the periodic
geometry were extended outward to a distance of about $10^{22}$
light years, then the energy of the radiation would dominate that
of the source; and, even at linearized order and with ignorance
of the time evolution, the periodic geometry would no longer
resemble the physically realistic one.
For this system, only at such a large distance does the
lack of asymptotic flatness of the periodic geometry intrude
upon the analysis.

A second example is provided by a test particle of small mass
$\mu$ close to the innermost stable circular orbit about a black
hole of mass $M$ with $\mu \ll M$.  A perturbation analysis,
shows that the test particle nearly moves along a geodesic; the
secular corrections are from radiation reaction effects at order
$(\mu / M)^2$.  The physically reasonable solution to this system
involves only outgoing radiation.  But a corresponding exact
periodic solution can be constructed by demanding standing wave
boundary conditions at a large but finite distance from the black
hole.  The energy density of the gravitational waves at a
distance $r$ is proportional to $(\mu / M r)^2$.  As long as the
radius $r$ of the boundary is large enough that $r \gg M$ but
small enough that $r(\mu/M)^2 \ll M$ then the energy content of
the radiation will be dwarfed by the mass of the black hole, and
the boundary will still be in the weak-field zone.

These periodic solutions of the Einstein
equations are interesting, involve strong gravitational
fields, high speeds, are not asymptotically flat and cannot be
described by the linearized Einstein equations except in the
weak-field zone.  These are typical of the solutions of interest
in this paper.

The relationship between periodic geometries and physically
realistic ones with outgoing waves was carefully considered in
Paper~I\cite{paperI}.  There it was shown that a specific linear
combination of similar periodic geometries, of differing
frequencies with each geometry containing only standing waves in
the weak-field zone, was an approximation to an exact solution of
the Einstein equations with outgoing waves. Similar methods were
first used in the context of general relativity by
Thorne\cite{ThorneIII} who analyzed perturbations of neutron star
models.

The error in this approximation was also carefully analyzed in
Paper I.  When the effect of radiation reaction is weak, the
linear combination is sharply peaked at a resonant frequency.
And some physical quantity, describing an aspect of the linear
combination of metrics, differs from the corresponding quantity
for the exact outgoing-wave metric by an amount comparable to its
change in one cycle due to the effects of radiation reaction.
Thus, a resonant periodic solution alone accurately models an
outgoing-wave solution as long as the radiation reaction
timescale is much longer than the dynamical timescale.  And the
error in the linear combination is inversely proportional to the
ratio of these two timescales; this is also proportional to the
ratio of the frequency width of the linear combination to the
resonant frequency.

Relativistic binary systems are, perhaps, the most interesting of
the systems which can be approximated in this way. In particular
the gravitational wave luminosity and the location of the
innermost stable orbit in a relativistic, binary system can be
studied. Other phenomena of interest which are approachable via
periodic geometries include axisymmetric, rapidly rotating
neutron star models and quasi-normal oscillations of black holes
or neutron stars. However, even in circumstances when radiation
reaction forces are strong and evolution is normally rapid, the
periodic solutions are still interesting in their own
right---after all they are solutions of the Einstein equations
with generally strong gravitational fields and gravitational
radiation.

Thus far the two most useful methods of studying relativistic
binaries are the post-Newtonian approximation and the test
particle approximation.  In both of these approaches the effects
of radiation reaction are ignored at the first order of a small
quantity, and the consequent orbits studied are conservative and
periodic. The periodic assumption includes both of these
approaches as special cases and is, therefore, of more general
validity.

Paper I contains a variational principle restricted to time
independent geometries; in this paper the variational principle is
generalized to include periodic geometries. The generalization
reveals the role of gravitational radiation more clearly and
substantially simplifies the treatment of the radiative boundary
terms in the weak-field zone.  Also, in Paper I the boundary
terms had to be evaluated in the local wave zone; in this paper
it is necessary to go out only to the weak-field zone where the
geometry is accurately approximated by the linearized Einstein
equations---for the Sun-Earth system this is the difference
between going out to Alpha Centauri for the local wave zone and
out to the orbit of Jupiter for the weak field zone.  And given
an approximate periodic geometry, which differs from an exact
solution to the Einstein equations by order $\delta$, the
variational principle has the power to yield accurate estimates,
with an error of order $\delta^2$, of {\em all} the interesting
quantities which describe the geometry, except the actual metric
itself.  These accurately estimated quantities include the
effective mass and angular momentum of the total system as well
as its individual components, the frequency of periodicity, the
angular frequency of rotation, and the amplitudes and phases of
the gravitational radiation in a multipole decomposition.

The quantity being extremized in the variational principle is
called the effective mass of the system.  It is defined as a
surface integral in the weak-field zone, and has the interesting
property of being independent of the actual surface chosen for
the integral, through terms of second order in the deviation of
the metric from flat space as long as the surface is in the
weak-field zone.  In particular, the effective mass does not
include the mass content of the standing gravitational waves,
which diverges and keeps the geometry from being
asymptotically flat.  And a second surface integral, involved in
the variational principle, defines the effective angular momentum
in a similar manner which also does not include a divergent
contribution from the standing gravitational waves.

Section \ref{background} first reviews the 3+1 formalism of
general relativity and, then, reviews and modestly modifies
Thorne's\cite{ThorneRMP} unique notation for symmetric trace free
tensors, outlines his general solution of the linearized Einstein
equations, and transforms to a gauge more suitable for present
purposes. Previous familiarity with Thorne's notation greatly
facilitates the reading of this paper. Section \ref{VP} gives a
variational principle for a generic, periodic solution to the
Einstein equations.  The actual quantity being extremized is
closely related to the quasi-local energy as developed by Brown
and York\cite{YorkQLE}, and the appropriate boundary conditions
for the variational principle include the specification of the
three metric on the sides of the bounded region of space-time
under consideration.  In \S\ref{VPweak} the variational principle
is modified for the special case that the boundary is in the
weak-field zone, and the boundary integrals are rewritten in
terms of the amplitude and phase of the gravitational waves of
different multipoles.  The variational principle leads to formal
definitions of the effective mass and effective angular momentum
of a periodic geometry.  Some of the complicated analyses are
described in the Appendix.

\section{Background and Notation} \label{background}
\subsection{The Initial Value Formalism and
      Dynamics of General Relativity}

A four dimensional space-time with a metric \(g_{ab} \) may be
foliated into spacelike hypersurfaces of constant \(t\), with a
metric \( \gamma_{ab}\),\cite{YorkIV}
\begin{equation}
   g_{ab}\,dx^a dx^b = -N^2 dt^2
   + \gamma_{ab}(dx^a + N^a dt)(dx^b + N^b dt).
\end{equation}
The quantity \(N\) is the lapse function, and
\(N^a\) is the shift vector. The three dimensional metric has a
derivative operator \(D_a \), and Ricci tensor \(R_{ab} \).

The extrinsic curvature, \(K_{ab} \), of the hypersurface is
defined from
\begin{eqnarray}
      {\cal G}_{ab} & \equiv &  - 2 N K_{ab}
         + 2 D_{\left(a\right.} N_{\left.b\right)}
          = {\cal L}_t \gamma_{ab},
\label{LieGamma}
\end{eqnarray}
where \( {\cal L}_t \) is the Lie derivative with respect to the
time translation vector,
\(t^a \partial / \partial x^a \equiv \partial / \partial t \),
which points in the direction of increasing \(t\) with all
spatial coordinates held fixed.
In the Hamiltonian formulation of general relativity\cite{ADM},
the
momentum conjugate to \( \gamma_{ab} \) is \( \pi^{ab}/16\pi \)
where
\begin{equation}
      \pi^{ab} \equiv - \gamma^{1/2} \left(K^{ab} - \gamma^{ab}
      {K^c}_c\right),
\label{defPi}
\end{equation}
and $\gamma$ is the determinant of $\gamma_{ab}$.

The constraint equations on a given hypersurface are restrictions
on \( \gamma_{ab} \) and \( \pi^{ab} \)
from the Einstein equations.  These are the Hamiltonian
constraint,
\begin{equation}
        {\cal N} \equiv R + \gamma^{-1} (\frac{1}{2}
                        {\pi_a}^a {\pi_b}^b
                        - \pi^{ab} \pi_{ab}) = 16 \pi \rho,
\label{HamConstr}
\end{equation}
and the momentum constraint,
\begin{equation}
     {\cal N}^a \equiv D_b(\pi^{ab} /\gamma^{1/2}) = -8\pi j^a,
\label{MomConstr}
\end{equation}
where \(\rho\) is the energy density and \(j^a \) is the
momentum density of the stress energy of matter,
\begin{eqnarray}
   \rho & = & N^2 T^{tt},   \\
   j^a & =  & N \gamma^{ab} {T_{b}}^t,
\end{eqnarray}
and the spatial part of the stress-energy tensor,
\begin{equation}
   S_{ab} \equiv {\gamma_a}^c {\gamma_b}^d T_{cd},
\end{equation}
is used below.

The dynamical part of the Einstein equations
gives
\begin{eqnarray}
{\cal P}^{ab} & \equiv &
   -N\gamma^{1/2}\left(R^{ab}- \frac{1}{2} \gamma^{ab}R\right)
   +\frac{1}{2} N\gamma^{-1/2}\gamma^{ab}
               \left(\pi^{cd}\pi_{cd} - \frac{1}{2}
                           {\pi^c}_c {\pi^d}_d\right)
\nonumber\\&&
   -2N\gamma^{-1/2}
        \left(\pi^{ac}\pi_{c}{}^{b}
            -\frac{1}{2}{\pi^c}_c \pi^{ab}\right)
   +\gamma^{1/2}\left(D^aD^bN-\gamma^{ab}D^cD_cN\right)
\nonumber\\&&
   + \gamma^{1/2} D_c \left( \gamma^{-1/2}\pi^{ab}N^c \right)
   - \pi^{ac} D_c N^b - \pi^{bc} D_c N^a
   + 8 \pi N \gamma^{1/2} \left( S^{ab}
         - \rho \gamma^{ab} \right)
\nonumber \\
   & = & {\cal L}_t \pi^{ab}.
\label{LiePi}
\end{eqnarray}

\subsection{The Linearized Analysis in the Weak-Field Zone}
\label{weakfield}

The space-times of interest in this paper are periodic
and may contain gravitational standing waves far from the source.
And periodic, standing waves cannot extend to spatial infinity
in an asymptotically flat geometry\cite{gibbons84}.
However, if the amplitude of the waves is small enough then a
weak-field zone around the source exists wherein the geometry is
well approximated by the linearized Einstein equations.  It is in
this weak-field zone that we analyze the gravitational waves and
define quantities similar to the mass and angular momentum of the
system.

\subsubsection{Notation}
\label{Notation}
For the geometry in the weak-field zone, a slight
modification is made of Thorne's\cite{ThorneRMP} notation for
coordinates and symmetric trace-free tensors.
The lapse function at zeroth order in the deviation from flat
space is a constant $N_0$, not necessarily
unity; thus the Minkowskii time coordinate is $N_0t$, and the
Cartesian coordinates \( (x,y,z) \) are the traditional flat
space coordinates, with \((r,\theta,\phi)\) being the
corresponding spherical coordinates. Indices \( (i,j,k,p,q,r,s)
\) run over \((x,y,z)\) and denote the Cartesian components of a
spatial tensor with a flat metric. Summation is implied when
these indices are repeated, and such indices are sometimes both
lowered without ambiguity. The flat space derivative operator
is~\(\nabla_i\). A comma denotes a partial derivative with
respect to a Cartesian coordinate.  And the quantity
\(r_i\equiv\nabla_ir\) is the unit outward radial vector.

Tensors with large numbers (say \(l\)) of indices are common.
Thus in the convenient abbreviation
\begin{equation}
      {T}_{J_l} \equiv {T}_{j_1 j_2 \ldots j_l}
\end{equation}
the indices represent tensor components in a Cartesian coordinate
system, just like \((i,j,k,p,q,r,s)\) do. Furthermore, often such
tensors are both symmetric and trace-free (STF) on all pairs of
indices
and also have Cartesian components which are functions only of
\(t\) and \(r\) and independent of \(\theta\) and \(\phi\); the
Cartesian components of such STF tensors are written as
capital script letters. For example,
\begin{equation}
      {\cal T}_{J_l} = {\cal T}_{j_1 j_2 \ldots j_l}
      = {\cal T}_{\left(j_1 j_2 \ldots j_l\right)},
\end{equation}
\begin{equation}
      {\cal T}_{j_1 j_1 j_3 \dots j_l} = 0
\end{equation}
and
\begin{equation}
   \partial {\cal T}_{J_l} / \partial \theta  =  0 ,\;\;
   \partial {\cal T}_{J_l} / \partial \phi    =  0 .
\end{equation}
Also the abbreviation
\begin{equation}
   R_{K_l} \equiv r_{k_1} r_{k_2} \ldots r_{k_l}
\end{equation}
is useful to represent the outer product of many unit radial
vectors.

A convenient decomposition of the STF $l$-tensors is in terms of
a basis set of \(2l+1\) constant STF
tensors~\( {\cal Y}^{lm}_{J_l}\), with \(-l \leq m \leq l \),
defined by Thorne (his Eq.~(2.12)).  Two useful properties of
the \( {\cal Y}^{lm}_{J_l} \) are
\begin{equation}
   Y^{lm} = {\cal Y}^{lm}_{J_l} R_{J_l},
\end{equation}
where \(Y^{lm}\) is the usual spherical harmonic function, and
their orthogonality
\begin{equation}
   {\cal Y}^{lm}_{J_l} {\cal Y}^{lm^{\prime}\ast}_{J_l}
      = \frac{(2l+1)!!}{4\pi l!}
            \delta^{mm^{\prime}}.
\end{equation}
With this basis set of STF tensors the decomposition of
\( {\cal T}_{J_l}(t,r) \) yields
\begin{equation}
   {\cal T}_{J_l}(t,r) = \sum_{m=-l}^{l} T_{lm}(t,r)
                                       {\cal Y}^{lm}_{J_l}.
\end{equation}

Also if \( {\cal T}_{J_l}(t,r)/r \) is a solution to the flat
space wave equation then it may be separated into outgoing and
ingoing parts,
\begin{equation}
   {\cal T}_{J_l}(t,r) = {\cal T}_{J_l}^{\text{out}}(N_0 t - r)
                      + {\cal T}_{J_l}^{\text{in}}(N_0 t + r) .
\end{equation}
And the decompositions,
\begin{eqnarray}
   {\cal T}_{J_l}^{\text{out}}(N_0t-r) & = & \sum_{m=-l}^{l}
               T_{lm}^{\text{out}}(N_0t-r) {\cal Y}^{lm}_{J_l},
\\
   {\cal T}_{J_l}^{\text{in}} (N_0t+r) & = & \sum_{m=-l}^{l}
                T_{lm}^{\text{in}}(N_0t+r) {\cal Y}^{lm}_{J_l},
\end{eqnarray}
are natural.
Furthermore, if $T_{lm}^{\text{out}}$ and $T_{lm}^{\text{in}}$
are
each composed of a sum of periodic pieces of different
frequencies $\omega_{mn}$, then complex amplitudes $T^{n}_{lm}$
and phases $\theta^{Tn}_{lm}$ are defined from
\begin{equation}
   T_{lm}^{\text{out}}(N_0t-r)  =  \sum_{n} T^{n}_{lm}
     \exp [-i\theta^{Tn}_{lm} + \frac{1}{2} il\pi +
            i\omega_{mn}(N_0t-r)],
\label{defTphase1}
\end{equation}
and
\begin{equation}
   T_{lm}^{\text{in}}(N_0t+r)  =  \sum_{n} T^{n}_{lm}
     \exp [ i\theta^{Tn}_{lm} - \frac{1}{2} il\pi
         + i \omega_{mn}(N_0t+r)] .
\label{defTphase}
\end{equation}
If the waves are standing, then the $\theta^{Tn}_{lm}$ are
real so that the outgoing and ingoing magnitudes are equal.
The above equations lead to
\begin{equation}
   T_{lm}(t,r) = \sum_n 2 T^n_{lm}
         \cos(\theta^{Tn}_{lm} - \frac{1}{2} l\pi
         + \omega_{mn} r)
         e^{i\omega_{mn}N_0t}
\end{equation}
and
\begin{equation}
   {\cal T}_{J_l}(t,r) = \sum_n \sum_{m=-l}^l
         2 T^n_{lm} {\cal Y}^{lm}_{J_l}
         \cos(\theta^{Tn}_{lm}
               - \frac{1}{2} l\pi + \omega_{mn} r)
         e^{i\omega_{mn}N_0t}.
\end{equation}

Thorne defines ${\cal Y}^{lm}_{J_l}$ as a special STF
tensor which depends upon a particular orientation of the
Cartesian axes.  Thus one frame of reference rotated with respect
to a second has a different corresponding
${\cal Y}^{lm}_{J_l}$.  For a frame of
reference rotating with a coordinate angular
velocity $\Omega$ about the
$z$-axis with respect to an inertial frame
\begin{equation}
   {{\cal Y}^{lm}_{J_l}}_{\text{rot}} = e^{-im\Omega t}
                        {{\cal Y}^{lm}_{J_l}}_{\text{inertial}},
\end{equation}
from Thorne's
Eq.~(2.12) if the axes are aligned at $t=0$.
And as viewed from the rotating frame
\begin{equation}
   {\cal T}_{J_l}(t,r) = \sum_n \sum_{m=-l}^l
         2 T^n_{lm} {{\cal Y}^{lm}_{J_l}}_{\text{rot}}
         \cos(\theta^{Tn}_{lm}
               - \frac{1}{2} l\pi + \omega_{mn} r)
         e^{i(\omega_{mn}+m\Omega/N_0)N_0t}.
\label{T-rot}
\end{equation}

The final specialization of interest
is that ${\cal T}_{J_l}$ be periodic with fundamental frequency
$\omega_0$ as viewed from the rotating frame of reference.
Eq.~(\ref{T-rot}) then requires that
\begin{equation}
   \omega_{mn} = n \omega_0 - m \Omega / N_0,
\end{equation}
for $n$ an integer.

If ${\cal T}_{J_l}$ is constant in time in the inertial
frame of reference then $\omega_{mn}$ must vanish for all $m$ and
$n$ for which $T^{n}_{lm}$ is not zero.
If, in addition, $\Omega$ is nonzero and $\Omega/\omega_{0}N_{0}$ is not
rational then the constant $T^{n}_{lm} = 0$ unless $m=n=0$, i.e.
${\cal T}_{J_l}$ is constant in time and axisymmetric;
or if $\Omega$ is zero then $T^{n}_{lm}=0$ for $n \neq 0$.
Also, if $\Omega/\omega_{0}N_{0}$ is an integer then
${\cal T}_{J_l}$
is also periodic when viewed from an inertial frame of reference.

{}From Thorne's definition of
\( {\cal Y}^{lm}_{J_l} \) it follows that
\begin{equation}
      {\cal Y}^{lm}_{J_l} = (-1)^m {\cal Y}^{l,-m\ast}_{J_l},
\end{equation}
and consequently if ${\cal T}_{J_l}$ is real then
\begin{equation}
         \omega_{m,n} = - \omega_{-m,-n},
\end{equation}
\begin{equation}
      \theta_{l,m}^{T,n} = - \theta_{l,-m}^{T,-n},
\end{equation}
and
\begin{equation}
      T^n_{l,m} = (-1)^{l+m} T_{l,-m}^{-n\ast}.
\end{equation}

\subsubsection{General Linearized Solution}

Thorne\cite{ThorneRMP} gives a general solution to the
linearized Einstein equations
in terms of the mass and current moments evaluated at a retarded
time, $ {\cal I}_{K_l}(N_0 t - r)$
and ${\cal S}_{K_l}(N_0 t - r)$, in one specialization of the
Lorentz gauge.
It is
convenient to separate the stationary, time independent parts of
the metric from the (often radiative) time dependent parts.
The superscript ${}^0$ refers to the constant moments
so that \({\cal I}^0 \) and
\({\cal S}^0_j \) are the constant mass monopole and
current dipole moment of the geometry---the mass and
angular momentum of the geometry if there is no gravitational
radiation.
Useful definitions are
\begin{eqnarray}
   I & \equiv & \frac{2 {\cal I}^0}{r}
      + \sum_{l=2}^{\infty} \frac{2(2l-1)!!}{r^{l+1}l!}
      {\cal I}^0_{K_l} R_{K_l}
\label{defI}
\end{eqnarray}
and
\begin{eqnarray}
   S^j & \equiv & \frac{-2 f^{ji} \bbox{\epsilon}_{ipq}
                                 {\cal S}_p^0 r_q}{r^2}
      - f^{ji} \sum_{l=2}^{\infty}
      \frac{4l(2l-1)!!}{r^{l+1}(l+1)!} \bbox{\epsilon}_{ipq}
         {\cal S}^0_{pK_{l-1}} r_q R_{K_{l-1}}.
\label{defSj}
\end{eqnarray}
The boldface \(\bbox{\epsilon}_{jpq} \) is the Levi-Civita
tensor.

Thorne's metric is just a perturbation away from flat space-time,
and
consequences of his Eq.~(8.13),
through first order in the perturbation, are
\begin{equation}
  N = N_0 \left\{ 1 - \frac{1}{2} I -  \sum_{l=2}^{\infty}
    \frac{\left(-\epsilon\right)^l}{l!}
    \left[r^{-1} {\cal I}_{K_l}
   (N_0t-\epsilon r)\right],_{K_l} \right\}
\label{defNa}
\end{equation}
and
\begin{eqnarray}
   N^j & = & N_0 S^j
     + N_0 f^{ji} \sum_{l=2}^{\infty} \frac{4l(-\epsilon)^{l+1}}{(l+1)!}
      \left[r^{-1} \bbox{\epsilon}_{ipq}
      {\cal S}_{pK_{l-1}}(N_0t-\epsilon r)\right],_{qK_{l-1}}
\nonumber \\ & &
      +N_0 f^{ji} \sum_{l=2}^{\infty}
   \frac{4(-\epsilon)^l }{l!} [ r^{-1}
      \dot{\cal I}_{iK_{l-1}} (N_0t-\epsilon r) ],_{K_{l-1}}
\end{eqnarray}
for the lapse and shift, a dot represents a derivative with
respect to the function argument \((N_0t-\epsilon r)\) and
\(\epsilon\) is $+1$ for outgoing radiation and $-1$ for ingoing.
The three-metric, through first order,
has the stationary part separated from the time dependent part by
\begin{eqnarray}
\gamma_{ij}  & = & f_{ij} (1+I) + h_{ij}
\label{gammaij}
\end{eqnarray}
where $f_{ij}$ is the flat three metric in Cartesian components,
and the time dependent terms are all contained in
\begin{eqnarray}
   h_{ij} & = & f_{ij}
   \sum_{l=2}^{\infty} \frac{2(-\epsilon)^l}{l!}
   \left[r^{-1} {\cal I}_{K_l} (N_0t-\epsilon r)\right],_{K_l}
\nonumber\\&&
   + \sum_{l=2}^{\infty} \frac{4 (-\epsilon)^l}{l!}
   \left[r^{-1} \ddot{\cal I}_{ijK_{l-2}}
               \left(N_0t-\epsilon r\right)\right],_{K_{l-2}}
\nonumber\\&&
   + \sum_{l=2}^{\infty} \frac{8l(-\epsilon)^{l+1}}{(l+1)!}
      \left[r^{-1} \bbox{\epsilon}_{pq\left(i\right.}
      \dot{\cal S}_{\left.j\right)pK_{l-2}}
              \left(N_0t-\epsilon r\right)\right],_{qK_{l-2}}.
\label{hija}
\end{eqnarray}

A different version of the Lorentz gauge is preferred here
wherein all of
the time dependence is removed from the lapse and shift vector.
The gauge change generated by \((\xi_t, \xi_i) \),
where
\begin{eqnarray}
   \partial\xi_t / \partial t & = &
      - N_0^2 \sum_{l=2}^{\infty}
      {(-\epsilon)^l \over l!} \left[r^{-1}{\cal I}_{K_l}
      (N_0t-\epsilon r)\right],_{K_l}
\nonumber \\
   \partial\xi_j / \partial t & = &  - \nabla_j \xi_t
                                          - (N_j - N_0S_j),
\end{eqnarray}
accomplishes this.
The ultimate result is that the general solution to the
linearized Einstein equations, in the preferred gauge,
has a lapse and shift vector
\begin{equation}
   N = N_0 (1-I/2),
\label{N}
\end{equation}
and
\begin{equation}
   N^j = N_0 S^j.
\label{Nj}
\end{equation}
And $\gamma_{ij}$ is still given by
Eq.~(\ref{gammaij}) but with
\begin{eqnarray}
   h_{ij} & = & \sum_{l=2}^{\infty} (-\epsilon)^l \frac{2}{l!}
   \left\{
      f_{ij} [r^{-1} {\cal I}_{K_l}],_{K_l}
      + 2 [r^{-1}
            {\cal I}^{\left(+2\right)}_{ijK_{l-2}} ],_{K_{l-2}}
      + [r^{-1}{\cal I}^{\left(-2\right)}_{K_{l}} ],_{ijK_{l}}
      - 4 [r^{-1}{\cal I}_{K_{l-1}(i} ],_{j)K_{l-1}}
   \right\}
\nonumber\\&&
      + \sum_{l=2}^{\infty} \frac{(-\epsilon)^{l+1} 8l}{(l+1)!}
      \left\{
         [r^{-1} \bbox{\epsilon}_{pq\left(i\right.}
            {\cal S}^{\left(+1\right)}_{\left.j\right)pK_{l-2}}
                                                 ],_{qK_{l-2}}
         - [r^{-1} {\cal S}^{\left(-1\right)}_{pK_{l-1}}
            \bbox{\epsilon}_{pq\left(i\right.}
                                    ],_{\left.j\right)qK_{l-1}}
      \right\} .
\label{hijb}
\end{eqnarray}
where the parenthesized superscript,
e.g. \({\cal S}^{\left(-1\right)}_{pK_{l-1}}\),
denotes differentiation, with respect to the implied argument
\( (N_0 t - \epsilon r) \) the appropriate number of times,
positive or negative.
Also
\begin{equation}
   K_{ij} = - \frac{1}{2N_0}\frac{\partial h_{ij}}{\partial t}
                  + \nabla_{(i} S_{j)}.
\label{Kij}
\end{equation}

In this gauge a number of useful identities hold:
\begin{equation}
   f^{ij} h_{ij} = 0,
\label{hii}
\end{equation}
\begin{equation}
   f^{ij} K_{ij} = 0,
\label{kii}
\end{equation}
\begin{equation}
   \nabla_{i} {h^i}_j = 0,
\label{Dihij}
\end{equation}
\begin{equation}
   r_{i} S^{i} = 0,
\label{riSi}
\end{equation}
\begin{equation}
   \nabla_{i} S^{i} = 0,
\label{DiSi}
\end{equation}
\begin{equation}
   \nabla_{k} \nabla^{k} I = 0,
\label{DkDkI}
\end{equation}
\begin{equation}
   \nabla_{k} \nabla^{k} S^{i} = 0,
\label{DkDkSi}
\end{equation}
and
\begin{equation}
   \nabla_{k} \nabla^{k} h_{ij}
      - \frac{1}{N_{0}^{2}}\frac{\partial^2 h_{ij}}{\partial t^2}
      = 0.
\label{DkDkhij}
\end{equation}

In the local wave zone the leading $1/r$ contributions
to $h_{ij}$ from the time dependent moments give
\begin{eqnarray}
   h_{ij} & = &
      \sum_{l=2}^{\infty} \frac{4}{rl!}
          {\cal I}^{\left(l\right)}_{pqK_{l-2}} R_{K_{l-2}}
         ({\sigma_0^p}_{(i}{\sigma_0^q}_{j)}
            - \frac{1}{2} \sigma_0^{pq} \sigma_{0ij})
\nonumber\\&&
    + \sum_{l=2}^{\infty} \frac{8l}{r(l+1)!}
      {\bf \epsilon}_{rsp} {\cal S}^{\left(l\right)}_{qrK_{l-2}}
      r_s R_{K_{l-2}}
         ({\sigma_0^p}_{(i}{\sigma_0^q}_{j)}
       - \frac{1}{2} \sigma_0^{pq} \sigma_{0ij}) + {\cal O}(r^{-2})
\end{eqnarray}
where $\sigma_{0ij}$ is the two dimensional metric of a
constant-$r$ two sphere.
Also in the wave zone
\(r^i h_{ij} = {\cal O}(r^{-2})\).

\subsubsection{The Standing Wave Solutions in the Rotating Frame
of Reference}
\label{StandingWave}

The amplitudes~$I^n_{lm}$ and phases~$\theta^{In}_{lm}$
of the gravitational radiation are defined in an inertial frame
of reference in a manner similar to Eqs.~(\ref{defTphase1})
and~(\ref{defTphase}):
\begin{equation}
   (-\epsilon)^l {\cal I}_{K_l}(N_0 t-\epsilon r) =
      \sum_{m,n} I^n_{lm} {\cal Y}^{lm}_{K_l}
      \exp [-i\epsilon(\theta^{In}_{lm}-\frac{1}{2} l \pi)
      + i\omega_{mn} (N_0t-\epsilon r)]
\end{equation}
and similarly for ${\cal S}_{K_l}$.

For the special case that the geometry is periodic in a rotating
frame of reference, and contains no traveling waves (only
standing waves) the notation of \S\ref{Notation} allows the
linearized geometry in the rotating frame of reference to
be written in terms of the stationary moments, $I^0_{lm}$
and $S^0_{lm}$, and the periodic, radiative moments $I^n_{lm}$
and $S^n_{lm}$ as
\begin{equation}
N = N_0 (1-I/2),
\label{defNb}
\end{equation}
and
\begin{equation}
N^j = N_0 S^j + \Omega \Phi^j
\label{defNj}
\end{equation}
where
\begin{equation}
   \Phi^i \partial / \partial x^i \equiv
                     \partial / \partial\phi ,
\label{defPhi}
\end{equation}
\begin{equation}
I = \frac{2 {\cal I}^0}{r} +
      \sum_{l,m}\frac{2(2l-1)!!}{r^{l+1}l!}
      I^0_{lm} {\cal Y}^{lm}_{K_l}R_{K_l},
\label{I}
\end{equation}
and
\begin{equation}
S^j =-\frac{2 f^{ji}\bbox{\epsilon}_{ipq} r_q {\cal S}_p^0}{r^2}
      - f^{ji} \sum_{l,m}
      \frac{4l(2l-1)!!}{r^{l+1}(l+1)!} \bbox{\epsilon}_{ipq}
       S^0_{lm} {\cal Y}^{lm}_{pK_{l-1}} r_q R_{K_{l-1}}.
\label{Sj}
\end{equation}
Also
\begin{eqnarray}
h_{ij} & = & \sum_{l,m,n} \frac{4}{l!} I^n_{lm}
                                 e^{in\omega_0 N_0 t}
   \left\{
      f_{ij}{\cal Y}^{lm}_{K_l}
         [r^{-1}\cos(\theta^{nI}_{lm}-\frac{1}{2} l\pi
             +\omega_{mn}r)],_{K_l}
   \right.
\nonumber\\&&
      - 2\omega_{mn}^2 {\cal Y}^{lm}_{ijK_{l-2}}
         [r^{-1}\cos(\theta^{nI}_{lm}
               - \frac{1}{2} l\pi+\omega_{mn}r)],_{K_{l-2}}
\nonumber\\&&
   \left.
      - \omega_{mn}^{-2} {\cal Y}^{lm}_{K_{l}}
         [r^{-1}\cos(\theta^{nI}_{lm}
            - \frac{1}{2} l\pi+\omega_{mn}r)
         ],_{ijK_{l}}
      - 4 {\cal Y}^{lm}_{K_{l-1}(i}
         [r^{-1}\cos(\theta^{nI}_{lm}
            - \frac{1}{2} l\pi+\omega_{mn}r)
         ],_{j)K_{l-1}}
   \right\}
\nonumber\\&&
   - \sum_{l,m,n} \frac{16l}{(l+1)!} S^n_{lm}
               e^{in\omega_0 N_0 t}
   \left\{
      \omega_{mn} \bbox{\epsilon}_{pq\left(i\right.}
      {\cal Y}^{lm}_{\left.j\right)pK_{l-2}}
      [r^{-1}\sin(\theta^{nS}_{lm}
         - \frac{1}{2} l\pi+\omega_{mn}r)
      ],_{qK_{l-2}}
   \right.
\nonumber\\&&
   \left.
      +\omega_{mn}^{-1} {\cal Y}^{lm}_{pK_{l-1}}
      \bbox{\epsilon}_{pq\left(i\right.}
      [r^{-1}\sin(\theta^{nS}_{lm}
         - \frac{1}{2} l\pi+\omega_{mn}r)
      ],_{\left.j\right)qK_{l-1}}
   \right\}.
\label{exphij}
\end{eqnarray}
In the wave zone this becomes
\begin{eqnarray}
   h_{ij} & = &
      \sum_{l,m,n} \frac{8(\omega_{mn})^l}{rl!}
      I^n_{lm} \cos(\theta^{nI}_{lm}+\omega_{mn}r)
      {\cal Y}^{lm}_{K_{l-2}pq} R_{K_{l-2}}
      ({\sigma_0^p}_i {\sigma_0^q}_j
         - \frac{1}{2}\sigma_0^{pq}\sigma_{0ij})
      e^{in\omega_0 N_0 t}
\nonumber \\
   && + \sum_{l,m,n} \frac{16l(\omega_{mn})^l}{r(l+1)!}
      S^n_{lm}
      \cos(\theta^{nS}_{lm}+\omega_{mn}r)
      \bbox{\epsilon}_{rsp}
      {\cal Y}^{lm}_{qrK_{l-2}} r^s R_{K_{l-2}}
      ({\sigma_0^p}_i {\sigma_0^q}_j -
            \frac{1}{2}\sigma_0^{pq}\sigma_{0ij})
      e^{in\omega_0 N_0 t}.
\end{eqnarray}
These equations give the general periodic solution to the
linearized Einstein equations, in the gauge described above,
with stationary moments, $I^0_{lm}$ and $S^0_{lm}$,
and constant wave amplitudes and phases,
\( (I^n_{lm}, \theta^{In}_{lm}) \) and
\( (S^n_{lm}, \theta^{Sn}_{lm}) \).
And for \(h_{ij}\) to represent real,
standing waves it is necessary that
\begin{equation}
I^n_{lm} = (-1)^{l+m} I^{-n\ast}_{l,-m},
\end{equation}
\begin{equation}
\theta^{In}_{lm} = - \theta^{I,-n}_{l,-m},
\end{equation}
and similarly for \( S^n_{lm} \) and \( \theta^{Sn}_{lm} \).

Finally, it is useful in \S\ref{VPweak} and in the Appendix
to separate out
the part of $h_{ij}$ in Eq.~(\ref{exphij})  which includes
contributions from both $I^n_{lm}$ and $S^n_{lm}$ and is summed
over $l$, with both $m$ and $n$ held fixed.  Thus
\begin{equation}
   h_{ij} = \sum_{m,n} h^{mn}_{ij},
\label{hmn}
\end{equation}
and
\begin{equation}
   \nabla_k \nabla^k h^{mn}_{ij} - \omega_{mn}^2 h^{mn}_{ij} = 0.
\label{hmnij}
\end{equation}


\section{A Variational Principle for Periodic Solutions of the
Einstein Equations}
\label{VP}
A variational principle comes from the traditional Hamiltonian
formalism of general relativity and is very closely related to
the Hamiltonian and the
concept of quasi-local energy as developed by
Brown and York\cite{YorkQLE}, their Eq.~(4.13).
The starting point is the definition
\begin{eqnarray}
   16 \pi H_1 & \equiv & - \int
      \left\{
         N \left[
            R + {1\over\gamma}
            \left(\frac{1}{2} {\pi_a}^a {\pi_b}^b
               - \pi^{ab}\pi_{ab}
            \right)
         \right]
       + 2 N^a D_b({\pi^b}_a/\gamma^{1/2})
      \right\} \sqrt{\gamma} \,d^3x
\nonumber \\&&
         + \oint_{r_B} 2 N^b \gamma^{-1/2}  {\pi_b}^a r_a
            \sqrt{\sigma}\,d^2x
         - \oint_{r_B} 2  N \sigma^{ab} D_{a}r_{b}
            \sqrt{\sigma} \,d^2x.
\label{defH1}
\end{eqnarray}
This volume integral is over a spacelike hypersurface of finite
extent which is bounded by a two-surface defined by a scalar
field, $r$, which is constant on the boundary, $r=r_B$.  Here in
\S\ref{VP} the vector
$r^a$ is the outward pointing unit normal to the bounding
two-surface, \(\sigma_{ab} \) is both the metric of the
two-surface and the
projection operator onto the two-surface,
\begin{equation}
   \sigma_{ab} \equiv
      \gamma_{ab} - r_{a} r_{b}, \label{defsigma}
\end{equation}
and $\sigma$ is its determinant; and there is no
restriction on the location of the boundary of the finite
region in which we are
interested---in particular it is necessary neither that the
boundary extend out to the weak-field zone nor that the scalar
field $r$ which defines the boundary be related to a flat space
radial coordinate.

An arbitrary, infinitesimal variation of \(N, N^a,
\gamma_{ab} \text{ and } \pi^{ab} \), which holds the location of
the boundary fixed, results in an infinitesimal change in
$H_1$,
\begin{eqnarray}
   16\pi \delta H_1 & = & - \int
      \left(
          \delta N {\cal N} \sqrt{\gamma}
         + 2 \delta N^a {\cal N}_{a} \sqrt{\gamma}
         + \delta \gamma_{ab} {\cal P}^{ab}
         - \delta \pi^{ab} {\cal G}_{ab}
      \right) \,d^3x
\nonumber \\ &&
      + \oint_{r_B} \left( r_a N^a
         \gamma^{-1/2} \pi^{bc} \delta\gamma_{bc}
         + 2 \delta N^b \gamma^{-1/2} {\pi_b}^a r_a
      \right) \sqrt{\sigma} \,d^2x
\nonumber \\ &&
      - \oint_{r_B} \left[
         N \delta \sigma^{ab} D_a r_b + ( 2 \delta N  +
                  N \sigma^{-1} \delta \sigma ) D_a r^a
         + \sigma^{-1} \delta \sigma r^a D_a N
      \right] \sqrt{\sigma} \,d^2x.
\label{deltaH1}
\end{eqnarray}
The indices of the perturbed quantities are neither raised nor
lowered by a metric. We now assume that a gauge may be chosen
with $r_a N^a = 0$ on the boundary.  Under some circumstances
this might not be possible as recently emphasized by
Hayward\cite{hayward93}; but for the applications which we
currently envision---namely a static boundary in the weak-field
zone---this is possible.  Thus the first term in the first
surface integral of Eq.~(\ref{deltaH1}) vanishes, and a
comparison of the coefficients of the variations of $N$, $N^a$,
$\gamma_{ab}$ and $\pi^{ab}$ with the Einstein
equations~(\ref{HamConstr}), (\ref{MomConstr}), (\ref{LiePi}),
and (\ref{LieGamma}), respectively, shows that $H_1$ is a
Hamiltonian for the vacuum Einstein equations when the boundary
conditions are that $N$, $N^a$ and $\sigma^{ab}$ are given
functions of time on the surface at $r_B$ so that the remaining
variations within the surface terms all vanish.

A tradition in introductory electricity and magnetism is
to consider solutions to Maxwell's equations inside a ``box'',
with the boundary conditions, say, of the potential being given
on the surface of the box.  The Hamiltonian $H_1$ and the
boundary conditions of the previous paragraph are appropriate for
the gravitational analog of this tradition.

We have particular interest in periodic geometries, when all of
$N$, $N^a$, $\gamma_{ab}$ and $\pi^{ab}$ are periodic in $t$ with
period $ T = 2\pi/ N_0\omega_0$.  It is most convenient to
consider the geometrical quantities to be functions of the
spatial coordinates and a dimensionless time coordinate $\tau
\equiv 2\pi t/T$; and $T$ is left as unknown and yet to be
determined.

Along with the periodicity, the definition of
the time average of a dynamical quantity, say $H_1$, is
\begin{eqnarray}
   \langle H_1 \rangle & \equiv & T^{-1} \int^T_0 H_1 \,dt
   = \frac{1}{2\pi} \int^{2\pi}_0 H_1 \,d\tau.
\end{eqnarray}

The action used in Maupertuis's principle
of least action plays an important role here.  This is referred
to as the M-action $A$ defined by
\begin{eqnarray}
   16\pi A & \equiv & \int^{T}_0
      \int \pi^{ab} {\cal L}_t \gamma_{ab} \,d^3x \,dt
\nonumber\\ & = &
   \int^{2\pi}_0 \int \pi^{ab} \partial\gamma_{ab} /
         \partial\tau \,d^3x\,d\tau .
\label{defA}
\end{eqnarray}

It follows from Eqs.~(\ref{deltaH1}) and~(\ref{defA}) that
\begin{eqnarray}
   16\pi \delta \langle H_1 \rangle & = & - \frac{1}{2\pi}
      \int_0^{2\pi} \int
      \biggl[
         \delta N {\cal N} \sqrt{\gamma}
         + 2 \delta N^a {\cal N}_{a} \sqrt{\gamma}
         - \delta \gamma_{ab}
            ({\cal L}_t \pi^{ab} - {\cal P}^{ab})
         + \delta \pi^{ab}
            ({\cal L}_t \gamma_{ab} - {\cal G}_{ab})
      \biggr] \,d^3x\,d\tau
\nonumber \\ &&
   + 16\pi \delta A / T
   + \frac{1}{2\pi} \int_0^{2\pi} \oint_{r_B}
      2 \delta N^b \gamma^{-1/2}{\pi^a}_b r_a
   \sqrt{\sigma} \,d^2x\,d\tau
\nonumber \\ &&
   - \frac{1}{2\pi} \int_0^{2\pi} \oint_{r_B} \left[
      N \delta \sigma^{ab} D_a r_b
      + (2 \delta N + N \sigma^{-1} \delta \sigma ) D_a r^a
      + \sigma^{-1} \delta \sigma r^a D_a N
   \right] \sqrt{\sigma} \,d^2x\,d\tau.
\label{deltaH1b}
\end{eqnarray}
Eq.~(\ref{deltaH1b}) reveals the use of $\langle H_1\rangle$
in a variational principle.
Consider the class of \( (N, N^a, \gamma_{ab}, \pi^{ab}) \)
restricted to periodic functions of $\tau$, with period $2\pi$,
with a specific, predetermined value for $A$ and which satisfy
boundary conditions at $r_B$ specified as particular periodic
functions of $\tau$ for $N$, $N^a$ and $\sigma^{ab}$.
Then \(\langle H_1 \rangle \), considered a functional of the
\( (N,N^a,\gamma_{ab},\pi^{ab}) \) in this class, is an
extremum under arbitrary infinitesimal variations of the
\( (N, N^a, \gamma_{ab}, \pi^{ab}) \), remaining in the class,
if and only if the
\( (N, N^a, \gamma_{ab}, \pi^{ab}) \) satisfy the Einstein
equations.

Furthermore, the unknown $T$ can be found from the variational
principle as well.  If only $A$ is changed by a small amount,
$\Delta A$, and the variational principle reapplied to find a
corresponding change $\Delta \langle H_1 \rangle $, then
Eq.~(\ref{deltaH1b}) shows that
\begin{equation}
T = \Delta A / \Delta \langle H_1 \rangle.
\end{equation}


\section{The Variational Principle Extended to the
      Weak-Field Zone} \label{VPweak}

The importance of boundary terms in the Hamiltonian of general
relativity was emphasized by Regge and
Teitelboim\cite{ReggeandT}---different boundary conditions
necessitate the inclusion of different boundary integrals in the
Hamiltonian. For data consisting of $N$, $N^a$ and $\sigma^{ab}$
on the boundary, $H_1$ is the appropriate Hamiltonian.
Instead the data could consist of, essentially, the derivatives
of some of these quantities normal to the boundary, in which case
the appropriate Hamiltonian would be that of Eq.~(\ref{defH1})
but without the boundary integrals at $r_B$. The variational
principle of \S\ref{VP} is closely related to an action
principle, which is usually the source of definitions for
quantities such as mass and angular momentum as particular
boundary integrals. Thus, for each different choice of a
description of the data at the boundary, similar analysis results
in a variational principle involving different appropriate
boundary integrals, different definitions of quantities similar
to the mass and angular momentum and a different $\langle H
\rangle$. For the periodic, radiative geometries of this paper
the amplitudes and phases of the multipole moments (rather
than the fields and their normal derivatives) are the convenient
independent boundary data.  And this choice determines the boundary
integrals of the variational principle and the quantities which
correspond to mass and angular momentum.

In this section we discuss quantities similar to the mass,
angular momentum and amplitude and phase of gravitational
radiation, all in the context of periodic solutions of the
Einstein equations.  Now, these quantities are generally not well
defined except at spatial or null infinity; and even there
technical difficulties persist except under the best of
circumstances.  In addition our analysis is limited to a bounded
region of space-time which specifically excludes any infinity.
It is not surprising then that our approach does not directly
involve an elegant, gauge invariant definition of any of these
quantities.

Thus we forsake explicit gauge invariance for a precise variational
principle involving the multipole moments of the geometry. While
the physical interpretation of these moments may be suspect,
they have the virtues of being precisely defined in the context
of the chosen gauge and straightforward to calculate.
Furthermore, after the fact, a periodic geometry can be analyzed
and if the amplitude of gravitational radiation (as defined in
this gauge dependent manner) is sufficiently small, then the
geometry near the boundary is of the form of the general
linearized solution of \S\ref{weakfield}. And the mass monopole,
${\cal I}^0$, and current dipole, ${\cal S}^0_j$ are reminiscent
of mass and angular momentum.

\subsection{Specifcation of Boundary Data}
\label{BoundaryData}

The first task of this section is to describe the data on the
boundary in terms of the amplitudes $(I^{n}_{lm},S^{n}_{lm})$ and
phases $(\theta^{In}_{lm},\theta^{Sn}_{lm})$ of the multipole
moments. The physical metric near the boundary is not necessarily
flat and may, in fact, be far from flat.  None the less, in the
vicinity of the boundary we sometimes use a spatial, Cartesian
coordinate system, ($x$, $y$, $z$) along with the Euclidean
tensors $h_{ij}$ and $S^i$, defined in Eqs.~(\ref{exphij}) and
(\ref{Sj}), in terms of the multipole moments; also, the
two-sphere at constant $r$ has a unit normal $r_0^i$ when
embedded in flat space and  $r^a$ when embedded in the geometry
described by the metric $\gamma_{ab}$.  The notation of
\S\ref{weakfield} is employed here except that the symbol $r$
used as a tensor index on one of the Cartesian tensors denotes
the implied contraction of the index with $r_0^i$.

The geometry is still assumed to be periodic, so all
tensors are periodic functions of the dimensionless time
coordinate $\tau$ with period $2\pi$.  Also the frame in
which the geometry is periodic is assumed to be uniformly
rotating with respect to the frame of reference tied to ($x$, $y$, $z$)
with angular velocity $\Omega$ about the
$z$-axis.

A prescription for uniquely specifying data on the boundary
follows: Choose $N_0$ and $r_B$, once and for all---these are
never changed.  Now choose values of $T$, $\Omega$, $I^{n}_{lm}$,
$\theta^{In}_{lm}$, $S^{n}_{lm}$, and $\theta^{Sn}_{lm}$. In
terms of these chosen values let each of the following
geometrical quantities have a value {\em exactly} equal to that
given by just the linearized theory as described in
Eqs.~(\ref{defNb}) to (\ref{exphij}):
\( \sqrt{\sigma} \),
\( N^2 \sqrt{\sigma} \),
\( \sqrt{\sigma}
\sigma^{ab} \),
\( N^a \),
\( N(\sigma_a^c \sigma_b^d D_{c} r_{d}
      - \frac{1}{2} \sigma_{ab} D_c r^c) \),
\( N^{-1} D_a r^a \),
\( r^a D_a N \) and
\( \gamma^{-1/2}{\pi_a}^b r_b \sqrt{\sigma} \).
Also, let \( S^a \) and \(\Phi^a\) represent the contravariant
tensors in generic coordinates whose components in the
Cartesian coordinates are given in Eqs.~(\ref{Sj}) and
(\ref{defPhi}) respectively. Note that each of these quantities
is independent of the others. For example, the determinant of
$\sqrt{\sigma} \sigma^{ab}$ is unity and does not depend upon
$\sqrt{\sigma}$.  And $N(\sigma_a^c \sigma_b^d D_{c} r_{d} -
\frac{1}{2} \sigma_{ab} D_c r^c)$ is trace free and independent
of $D_a r^a$.  These specific choices for data on the boundary
are made so that the boundary integrals in Eq.~(\ref{deltaH3b})
below are {\em precisely} equal to what would be expected through
second order in an expansion in powers of the deviation from flat
space.

\subsection{The Variational Principle}

The next step toward a variational principle is the
definition
\begin{eqnarray}
   16 \pi H_2 & \equiv & - \int \left\{
      N \left[
         R+{1\over\gamma} \left(
            \frac{1}{2} {\pi_a}^a {\pi_b}^b - \pi^{ab}\pi_{ab}
         \right)
      \right] + 2 N^a D_b({\pi^b}_a/\gamma^{1/2})
   \right\} \sqrt{\gamma} \,d^3x
\nonumber\\&&
   + \oint_{r_B} 2 N_0 S^a \gamma^{-1/2} {\pi_a}^b r_b
      \sqrt{\sigma}\,d^2x
   - \oint_{r_B} 2 N \sigma^{ab} D_{a}r_{b} \sqrt{\sigma} \,d^2x.
\label{defH3}
\end{eqnarray}

An arbitrary, infinitesimal variation of \(N,N^a,\gamma_{ab}%
\text{ and }\pi^{ab}\), which holds fixed the location of the
boundary, results in an infinitesimal change of
$\langle H_2 \rangle $ which is similar to the change in
$\langle H_1 \rangle $ in Eq.~(\ref{deltaH1b}) and is
simplified by the
substitutions of $A$, defined in Eq.~(\ref{defA}), and of a
quantity, $J$, which reduces to the
current dipole moment for the specific boundary data given;
\begin{eqnarray}
   8\pi J & \equiv &
      -\frac{1}{2\pi} \int^{2\pi}_0 \oint_{r_B}
        \Phi^a \gamma^{-1/2}
        {\pi_a}^b r_b \sqrt{\sigma} \,d^2x\,d\tau
\nonumber\\ & = & 8\pi {\cal S}^0_z .
\label{defJ}
\end{eqnarray}
Then
\begin{eqnarray}
   16\pi\delta \langle H_2 \rangle & = &
   - \frac{1}{2\pi} \int_0^{2\pi}
      \int [\text{E Eqns}] \,d^3x\,d\tau
   + 16\pi \delta A / T
   + 16\pi \Omega \delta J
\nonumber \\ &&
   + \frac{1}{\pi} \int_0^{2\pi} \oint_{r_B}
      \delta(N_0 S^a) \gamma^{-1/2}{\pi_a}^b r_b
   \sqrt{\sigma} \,d^2x\,d\tau
\nonumber \\ &&
   - \frac{1}{2\pi} \int_0^{2\pi} \oint_{r_B} \left[
      \delta (\sqrt{\sigma}\sigma^{ab})
         N (D_a r_b - \frac{1}{2} \sigma_{ab} D_c r^c)
      + \delta (N^2 \sqrt{\sigma}) N^{-1} D_a r^a
      \right.
\nonumber \\ &&
\left.  + 2 \delta(\sqrt{\sigma}) r^a D_a N
   \right]\,d^2x\,d\tau.
\label{deltaH3b}
\end{eqnarray}
The symbol \( [ \text{E Eqns} ] \) is an abbreviation for the
first integrand in Eq.~(\ref{deltaH1b}) and vanishes
when the Einstein equations are satisfied.

The two surface integrals in Eq.~(\ref{deltaH3b})
are examined in the Appendix where it is shown
that
\begin{eqnarray}
\lefteqn{
   \frac{1}{\pi} \int_0^{2\pi} \oint_{r_B}
         \delta(N_0 S^a) \gamma^{-1/2}{\pi_a}^b r_b
   \sqrt{\sigma} \,d^2x\,d\tau
}
\nonumber \\
\lefteqn{
   - \frac{1}{2\pi} \int_0^{2\pi} \oint_{r_B} \left[
      \delta (\sqrt{\sigma}\sigma^{ab})
         N (D_a r_b - \frac{1}{2} \sigma_{ab} D_c r^c)
      + \delta (N^2 \sqrt{\sigma}) N^{-1} D_a r^a
      + 2 \delta(\sqrt{\sigma}) r^a D_a N
   \right]\,d^2x\,d\tau.
}
\nonumber \\  & = &
  \frac{\delta N_0}{2\pi} \int_0^{2\pi} \oint_{r_B} [
      2 S^a \gamma^{-1/2} \pi_{ab} r^b
      - \frac{2N}{N_0} D_a r^a ] \sqrt{\sigma} \,d^2x\,d\tau
\nonumber \\  & &
   + \frac{N_0}{4\pi} \delta \left\{
      \int_0^{2\pi} \oint_{r_B}
      [ \frac{1}{2} h^{ij} r_0^k \nabla_k h_{ij} \right.
         -r_{0p} \nabla_i(h^{ij} \sigma_{0jq} h^{qp})
            + \frac{5}{2r} h_{rr} h_{rr}
            - \frac{1}{r} h_{ij} {h^i}_{k}\sigma_0^{jk}
      +
         I r_0^i \nabla_i I
\nonumber\\ & &
      +
         S^i S_i /r - S^i r_0^j \nabla_j S_i
      \left. +
            \frac{1}{2} r r_0^p \nabla_p (h_{ij})
                  r_0^q \nabla_q (h^{ij})
            - \frac{1}{2} h_{ij} r_0^p \nabla_p
                  (r r_0^q \nabla_q h^{ij})
         ] \sqrt{\sigma_0} \,d^2x\,d\tau
      \right\}
\nonumber\\&&
   - \frac{N_0}{8\pi} \sum_{m,n}
      (n \omega_0 - \frac{m \Omega}{N_0}) \delta \left\{
      \int_0^{2\pi}
      \oint_{r_B} \omega_{mn}^{-1}
         [
            r r_0^p \nabla_p (h^{mn}_{ij}) r_0^q \nabla_q
                           (h^{mn\ast}_{ij})
            - h^{mn\ast}_{ij} r_0^p \nabla_p
                           (r r_0^q \nabla_q h^{mn}_{ij})
         ]
      \sqrt{\sigma_0} \,d^2x\,d\tau \right\}
\nonumber\\&&
      + N_0 \sum_{lmn} \left[
         \frac{8 (\omega_{mn})^{2l+1}(l+1)(l+2)}
                                          {(l!)^2 l (l-1)}
            I^n_{lm} I^{n\ast}_{lm} \delta \theta^{In}_{lm}
         + \frac{32 (\omega_{mn})^{2l+1} l (l+1)(l+2)}
                                          {[(l+1)!]^2 (l-1)}
            S^n_{lm} S^{n\ast}_{lm} \delta \theta^{Sn}_{lm}
      \right].
\label{masterST}
\end{eqnarray}
The different terms on the right hand side of this equation come
from expressions in the Appendix.
Inside the total-$\delta$,
the first term is built into expression~(\ref{defST});
the next three are the second integral of
expression~(\ref{radST});
the three terms involving the stationary moments are
expressions~(\ref{deltaI}) and~(\ref{deltaS});
and the last two terms come from expression~(\ref{deltaomega}).
The part of Eq.~(\ref{masterST}) proportional to
$ (n \omega_0 - m \Omega/N_0) $ is
the second integral of expression~(\ref{deltaomega}).
And the last summation in Eq.~(\ref{masterST}), which is
proportional to the variation of the phases, is
expression~(\ref{deltatheta}).

Consideration of Eqs.~(\ref{deltaH3b}) and (\ref{masterST})
leads to natural re-definitions of $J$ and $A$ which
move the parts of
Eq.~(\ref{masterST}) proportional to $16\pi\Omega$ into
$\delta J$ and the parts proportional to
$8 N_0 \omega_0 ( = 16 \pi /T)$
into $\delta A$.
These changes dramatically simplify
Eq.~(\ref{deltaH3b}) to the result in Eq.~(\ref{deltaVP}) below.
Thus the
effective angular momentum, $J_\ast$, is defined by
\begin{eqnarray}
   8\pi J_\ast & \equiv &
      -\frac{1}{2\pi} \int^{2\pi}_0 \oint_{r_B}
        \Phi^b \gamma^{-1/2} {\pi_b}^a r_a
               \sqrt{\sigma} \,d^2x\,d\tau
\nonumber\\&&
   + \sum_{m,n}
     \frac{m}{8\omega_{mn}} \oint_{r_B}
     [ r r_0^p \nabla_p (h^{mn}_{ij}) r_0^q \nabla_q
                                (h^{mn\ast}_{ij})
      - h^{mn\ast}_{ij} r_0^p \nabla_p
                     (r r_0^q \nabla_q h^{mn}_{ij})
     ] \sqrt{\sigma_0} \,d^2x.
\label{defJstar}
\end{eqnarray}
And the effective M-action, $A_\ast$, is defined by
\begin{eqnarray}
   16\pi A_{\ast} & \equiv &
      \int^{2\pi}_0 \int \pi^{ab} \partial\gamma_{ab} /
            \partial\tau \,d^3x\,d\tau
\nonumber\\&&
   - \sum_{m,n}
      \frac{\pi n}{2\omega_{mn}} \oint_{r_B}
      [ r r_0^p \nabla_p (h^{mn}_{ij}) r_0^q \nabla_q
                                  (h^{mn\ast}_{ij})
         - h^{mn\ast}_{ij} r_0^p \nabla_p
                       (r r_0^q \nabla_q h^{mn}_{ij})
      ] \sqrt{\sigma_0} \,d^2x.
\label{defAstar}
\end{eqnarray}

Similarly it seems natural to redefine $\langle H_2 \rangle$ by
having it absorb the terms inside the total-$\delta$ in
Eq.~(\ref{masterST}).  But, instead, we introduce the effective
mass which is a boundary integral independent of $N_0$:
\begin{eqnarray}
   16 \pi  m_\ast & \equiv &
   - \frac{1}{2\pi} \int_0^{2\pi} \oint_{r_B}
      \frac{2N}{N_0} D_{a} r^{a}
   \sqrt{\sigma} \,d^2x\,d\tau
   + \frac{1}{2\pi} \int_0^{2\pi} \oint_{r_B}
      2 \nabla_{i} r_0^{i}
   \sqrt{\sigma_0} \,d^2x\,d\tau
\nonumber\\&&
   + \frac{1}{2\pi} \int_0^{2\pi} \oint_{r_B}
      2  S^a \gamma^{-1/2} {\pi_a}^b r_b
   \sqrt{\sigma} \,d^2x\,d\tau
\nonumber \\ &&
   - \frac{1}{4 \pi} \int_0^{2\pi} \oint_{r_B}
   \left[
      \frac{1}{2} h^{ij} r_0^k \nabla_k h_{ij}
      - r_0^p \nabla_i(h^{ij} \sigma_{0jq} {h^q}_p)
            + \frac{5}{2r} h_{rr} h_{rr}
            - \frac{1}{r} h_{ij} {h^i}_{k}\sigma_0^{jk}
   \right.
\nonumber\\&&
      + I r_0^i \nabla_i I
      + S^i S_i /r - S^i r_0^j \nabla_j  S_i
\nonumber\\&&
   \left.
      + \frac{1}{2} r r_0^p \nabla_p (h_{ij}) r_0^q \nabla_q (h^{ij})
      - \frac{1}{2} h_{ij} r_0^p \nabla_p
               (r r_0^q \nabla_q h^{ij})
   \right] \sqrt{\sigma_0} \,d^2x\,d\tau .
\label{defmstar}
\end{eqnarray}
The variation of the second integral is identically zero;
it is included so that $m_{\ast}$ has the expected value for,
say, the Schwarzschild geometry.

Now a variational principle for $m_\ast$ is based
upon
\begin{eqnarray}
   16 \pi N_0 m_\ast & = &
      - \frac{1}{2\pi} \int _0^{2\pi} \int \left\{
         N \left[ R+{1\over\gamma} \left(\frac{1}{2} {\pi_a}^a
                  {\pi_b}^b - \pi^{ab}\pi_{ab}\right)
            \right]
         + 2 N^a D_b({\pi^b}_a/\gamma^{1/2})
      \right\} \sqrt{\gamma} \,d^3x\,d\tau
\nonumber \\ &&
   + N_0 \times \text{[right hand side of Eq.~(\ref{defmstar})]}
\label{defVP}
\end{eqnarray}
And the variation of this equation yields
\begin{eqnarray}
   16\pi N_0 \delta m_\ast & = &
   \frac{1}{2\pi} \int_0^{2\pi}
      \int [\text{E Eqns}] \,d^3x\,d\tau
   + 16\pi \delta A_\ast / T
   + 16\pi \Omega \delta J_\ast
\nonumber\\&&
      + N_0 \sum_{lmn}
      \left[
         \frac{8 (\omega_{mn})^{2l+1}(l+1)(l+2)}
                                          {(l!)^2 l (l-1)}
            I^n_{lm} I^{n\ast}_{lm} \delta \theta^{In}_{lm}
      \right.
\nonumber\\&&
      \left.
         + \frac{32 (\omega_{mn})^{2l+1} l (l+1)(l+2)}
                                          {[(l+1)!]^2 (l-1)}
            S^n_{lm} S^{n\ast}_{lm} \delta \theta^{Sn}_{lm}
      \right] .
\label{deltaVP}
\end{eqnarray}

To apply the variational principle:
choose specific values for $J_\ast$, $A_\ast$ and each of the
$\theta^{In}_{lm}$ and $\theta^{Sn}_{lm}$, and consider the
class of periodic geometries described by
\( (N, N^a, \gamma_{ab}, \pi^{ab}) \) which have these specific
values. Eq.~(\ref{deltaVP}) shows that $m_\ast$, evaluated by
Eq.~(\ref{defVP}), is an extremum for a member of this class if
and only if the geometry is a solution of the Einstein equations.

As might be anticipated, the coefficients of
$\delta \theta^{In}_{lm} / 2\pi $ and
$\delta \theta^{Sn}_{lm} / 2\pi $ are just $ 16\pi $ times the
energy in one wavelength of the gravitational waves in the wave
zone as derived from Thorne's Eq.~($4.16'$)\cite{ThorneRMP}.

\subsection{The Interpretations of $A_{\ast}$, $J_\ast$,
   and $m_\ast$}

In this section we show that when the boundary is in the
weak-field zone, each of the effective quantities is
independent of the exact location of the boundary through terms
of second order in the deviation of the geometry from flat space.
Thus, for each of the equations of this section, equality is
understood to hold only through terms of second order.

The quantity $A_{\ast}$ is defined as a sum of a volume
integral and a surface integral in Eq.~(\ref{defAstar}).
And the difference in $A_{\ast}$ corresponding to two
different boundary surfaces, at $r_1$ and $r_2$, is
\begin{eqnarray}
 16 \pi \Delta A_{\ast} & \equiv &
               16\pi (A_{2\ast} - A_{1\ast}) =
      \int^{2\pi}_0 \int_{r_1}^{r_2} \pi^{ab} \partial\gamma_{ab} /
            \partial\tau \,d^3x\,d\tau
\nonumber\\&&
   - \sum_{m,n}
      \frac{\pi n}{2\omega_{mn}}
         \left( \oint_{r_2}-\oint_{r_1} \right)
      [ r r_0^p \nabla_p (h^{mn}_{ij}) r_0^q \nabla_q
                                  (h^{mn\ast}_{ij})
         - h^{mn\ast}_{ij} r_0^p \nabla_p
                       (r r_0^q \nabla_q h^{mn}_{ij})
      ] \sqrt{\sigma_0} \,d^2x .
\end{eqnarray}

Through terms of second order in the weak-field zone this is
\begin{eqnarray}
 16 \pi \Delta A_{\ast} & = &
    \sum_{m,n} \int_{r_1}^{r_2}
          {\pi n \omega_{mn}
          h^{mn}_{ij}h^{mn\ast}_{ij}
          \sqrt{\sigma_0}} \,d^3x
\nonumber\\&&
   - \sum_{m,n} \frac{\pi n}{2 \omega_{mn}}
       \int_{r_1}^{r_2} \nabla_p
      [ r  \nabla_p (h^{mn}_{ij}) r_0^q \nabla_q
                                  (h^{mn\ast}_{ij})
         - h^{mn\ast}_{ij}  \nabla_p
                       (r r_0^q \nabla_q h^{mn}_{ij})
      ] \sqrt{\sigma_0} \,d^3x .
\end{eqnarray}
And by use of the wave equation, Eq.~(\ref{hmnij}), for $h^{mn}_{ij}$
this simplifies to
\begin{eqnarray}
 16 \pi \Delta A_{\ast} & = &
    \sum_{m,n} \int_{r_1}^{r_2}
          \pi n \omega_{mn}
          h^{mn}_{ij}h^{mn\ast}_{ij}
          \sqrt{\sigma_0} \,d^3x
   - \sum_{m,n} \frac{\pi n}{2\omega_{mn}}
   \int_{r_1}^{r_2} 2
       \omega_{mn}^2 h^{mn}_{ij} h^{mn\ast}_{ij}
       \sqrt{\sigma_0} \,d^2x
\nonumber\\
   & = & 0.
\end{eqnarray}
Thus the effective M-action is independent of the surface over
which it is evaluated, through second
order in the deviation from flat space.

Similarly, the difference in effective angular momentum evaluated
at two different surfaces in the weak-field zone is
\begin{eqnarray}
   8\pi\Delta J_{\ast} &\equiv& 8\pi (J_{2\ast}-J_{1\ast}) =
      -\frac{1}{2\pi} \int^{2\pi}_0
   \left( \oint_{r_2}- \oint_{r_1} \right)
        \Phi^b \gamma^{-1/2} {\pi_b}^a r_a
               \sqrt{\sigma} \,d^2x\,d\tau
\nonumber\\&&
   + \sum_{m,n}
   \frac{m}{8\omega_{mn}}
   \left( \oint_{r_2}- \oint_{r_1} \right)
     [ r r_0^p \nabla_p (h^{mn}_{ij}) r_0^q \nabla_q
                                (h^{mn\ast}_{ij})
      - h^{mn\ast}_{ij} r_0^p \nabla_p
                     (r r_0^q \nabla_q h^{mn}_{ij})
     ] \sqrt{\sigma_0} \,d^2x.
\label{DeltaJast}
\end{eqnarray}
The first integral may be rewritten as
\begin{eqnarray}
\lefteqn{
   -\frac{1}{2\pi} \int^{2\pi}_0
   \int_{r_1}^{r_2}
          D_a ( \gamma^{-1/2} {\pi_b}^a \Phi^b )
              \sqrt{\gamma} \,d^3x\,d\tau =}
\hspace{1in}
\nonumber\\  && -\frac{1}{2\pi} \int^{2\pi}_0
   \int_{r_1}^{r_2}
    \left[ D_a(\gamma^{-1/2}{\pi_b}^a) \Phi^b
           + \gamma^{-1/2}\pi^{ab} D_a \Phi_b   \right]
           \sqrt{\gamma} \,d^3x\,d\tau.
\end{eqnarray}
The first term vanishes from the momentum constraint,
Eq.~(\ref{MomConstr}), and the second is easily  written in terms
of the deviation from flat space.  Also use of the wave equation,
Eq.~(\ref{hmnij}), for $h^{mn}_{ij}$ in Eq.~(\ref{DeltaJast})
results in
\begin{eqnarray}
   8\pi\Delta J_{\ast} & = &
      - \sum_{mn}
   \int_{r_1}^{r_2}
         \frac{1}{4} \omega_{mn} m
         h^{mn}_{ij} h^{mn\ast}_{ij}
    \sqrt{\sigma_0}\,d^3x
   + \sum_{m,n}
   \frac{m}{8\omega_{mn}}
   \int_{r_1}^{r_2}
     2  \omega_{mn}^2
         h^{mn}_{ij} h^{mn\ast}_{ij}
    \sqrt{\sigma_0} \,d^3x
\nonumber \\
      & = & 0.
\end{eqnarray}
Thus the effective angular momentum is independent of the surface
over which it is evaluated in the weak-field zone, through second
order in the deviation from flat space.  At first order
$J_{\ast}$ is just the first term in
Eq.~(\ref{defJstar}), which evaluates to ${\cal S}_z^0$, the
current dipole moment as defined by the multipole structure of
the geometry in the weak-field zone.  Thus we have the following
interpretation of the effective angular momentum:  In the
weak-field zone, at linear order, $J_{\ast}$ is the constant
current dipole moment, ${\cal S}_z^0$.  But when the second order
corrections are added to the linearized theory, then ${\cal
S}_z^0$, has a small contribution from the standing gravitational
waves which is linear in $r$. However, $J_\ast$ does not change
at second order.  Thus, through second order in the deviation
from flat space $J_\ast$ is the current dipole moment of the
source alone without a contribution from the standing
gravitational waves.

In a similar fashion the difference in effective mass evaluated
at two different surfaces, $\Delta m_{\ast}$, is defined from
Eq.~(\ref{defmstar}).

Now, $\Delta m_{\ast}$ is quadratic in the deviation of the
geometry from flat space---the linear contribution vanishes from
Eq.~(\ref{deltaVP}) and the fact that $\Delta m_{\ast}$ is zero
for flat space. It is straightforward but an exceedingly tedious
task to substitute the linearized geometry of \S\ref{weakfield}
into the definition of $\Delta m_{\ast}$. In the later stages the
details of the substitution and reduction are similar to that for
$\Delta A_{\ast}$ and $\Delta J_{\ast}$. And the result is that
$\Delta m_{\ast}$, too, vanishes through second order in the
deviation from flat space in the weak-field zone.

Thus the effective mass is independent of the location of the
surface over which it is evaluated in the weak-field zone through
second order in the deviation of the geometry from flat space.  At
first order the effective
mass is the first two terms in Eq.~(\ref{defmstar}) which
evaluates to ${\cal I}^0$, the mass monopole moment of the
linearized geometry of \S\ref{weakfield}.  Thus we have the
following intrepretation of the effective mass: In the weak-field
zone, at linear order, $m_{\ast}$ is the constant mass monopole
of the linearized geometry.  But, when the second order
corrections are included, then the mass
monopole has a small contribution, linear in $r$, from the
standing gravitational waves.  However, $m_{\ast}$ does not
change through second order in the deviation from flat space.
Thus, to the extent that it is possible to define such a
quantity, $m_{\ast}$ is the mass monopole of the source alone
without a contribution from the standing gravitational waves.

\section{Conclusions}
Paper I showed that the periodic solutions of the Einstein
equations, with standing waves, give valuable information about
physically realistic systems, with outgoing radiation. And the
variational principle of Eq.~(\ref{defVP}) is a valuable tool
for studying the periodic solutions.  In a preliminary
application in Paper I, the variational principle has been
relatively easy to apply in the study of close orbits of equal
mass black holes, without regard to the gravitational radiation.
The next paper in this series will use more sophisticated trial
geometries which form a complete set via an infinite sequence of
possible parameters. In particular these geometries will clearly
contain gravitational radiation.

For specific values of $J_{\ast}$, $A_{\ast}$, $\theta^{In}_{lm}$
and $\theta^{Sn}_{lm}$ and an approximate solution to the
Einstein equations, accurate to order $\delta$, the variational
principle directly gives an estimate of $m_{\ast}$ which is
accurate to order $\delta^2$.  But the variational principle can
be used to estimate $\Omega$, $\omega_0$, $|I^{n}_{lm}|^2$ and
$|S^{n}_{lm}|^2$ to order $\delta^2$ as well.  For example, to
find an accurate estimate of $\Omega$ just apply the variational
principle a second time with a slightly different choice for the
effective angular momentum, say a value of
$J_{\ast} + \Delta J_{\ast}$. The resulting value of the
effective mass will change by
an amount, say, $\Delta m_{\ast}$; then Eq.~(\ref{deltaVP})
shows that
\begin{equation}
      \Omega = N_0 \Delta m_{\ast} / \Delta J_{\ast} + {\cal O}
      (\delta^2)
\end{equation}
accurately estimates $\Omega$.  The other interesting quantities
describing the geometry can be found similarly.

Perhaps the most interesting results of this paper are the
definitions of the effective quantities---Eq.~(\ref{defmstar})
for the effective mass, Eq.~(\ref{defJstar}) for the effective
angular momentum, and Eq.~(\ref{defAstar}) for the effective
M-action.  These definitions capture the concept of the mass
monopole, current dipole and M-action of the source without
contribution from the gravitational radiation in the weak-field
zone.  They are independent of the location of the boundary,
through second order in the deviations from flat space, as long
as the boundary is in the weak-field zone.  A similar analysis
has been made of a toy problem of a mass, confined to
oscillations along the $z$ axis under the influence of a generic
potential, and attached to a semi-infinitely long string
stretching out along the $x$ axis.  For that case the analogous
variational principle is for the average energy of the mass
alone---the energy of the string does not contribute.

To add matter to this analysis it is simple to include matter
terms in $H_2$.  And to allow for black
hole sources Eq.~(\ref{defVP}) for $m_{\ast}$ still provides
the variational principle.  But the volume integral has
boundaries in the weak-field zone surrounding the spatial
infinity within each black hole as discussed in Paper I.  Then
the variations bring in additional surface terms,
resembling the ones in Eq.~(\ref{deltaVP}), but evaluated at
these additional boundaries.  These new surface terms correspond
to the effective mass and angular momentum and the phase of the
radiation in the weak-field zone within the holes.

\acknowledgments
Part of this research was supported by the National Science
Foundation, grant 91-07007 with the University of Florida. And it
was completed when the author was visiting the Jet Propulsion
Laboratory, as a NASA/ASEE Summer Faculty Fellow, and enjoying
the support and hospitality of Frank Estabrook and Hugo
Wahlquist.


\appendix
\section{Boundary Integrals in the Weak-Field Zone}
\label{App}

The detailed analysis leading to Eq.~(\ref{masterST})
is relegated to this appendix.

The geometry on the boundary is described at the beginning of
\S\ref{BoundaryData}.  In particular on the boundary a number of
quantities are exactly equal to their linearized counterparts.
Thus, just on the boundary 
\begin{equation}
   \sqrt{\sigma} = \sqrt{\sigma_0}(1+I-\frac{1}{2}h_{rr}),
\label{defsig}
\end{equation}
\begin{equation}
   N^2 \sqrt{\sigma} = N_0^2\sqrt{\sigma_0} (1-\frac{1}{2}h_{rr}),
\end{equation}
\begin{equation} \sqrt{\sigma} \sigma^{ab} = \sqrt{\sigma_0}
         [  \sigma_0^{ab} (1-\frac{1}{2}h_{rr}) -
                  \sigma_0^{ac}h_{cd}\sigma_0^{db} ] ,
\end{equation}
\begin{equation}
N^a = N_0 S^a + \Omega \Phi^a,
\end{equation}
\begin{eqnarray} 
   2N(\sigma_a^c \sigma_b^d D_{c} r_{d}
         - \frac{1}{2} \sigma_{ab} D_c r^c) & = & N_0 [
        - \frac{2}{r} \sigma_{0a}^c h_{cd} \sigma_{0b}^d
        + \frac{1}{r} \sigma_{0ab} \sigma_{0}^{cd} h_{cd}
        + \sigma_{0a}^{c} \sigma_{0b}^{d} r^e \nabla_e h_{cd}
  -\frac{1}{2} \sigma_{0ab} \sigma_{0}^{cd} r^e \nabla_e h_{cd}
\nonumber \\ &&
        - \sigma_{0a}^{c} \sigma_{0b}^{d} r^e \nabla_c h_{de}
        - \sigma_{0a}^{c} \sigma_{0b}^{d} r^e \nabla_d h_{ce}
        + \sigma_{0ab} \sigma_{0}^{cd} r^e \nabla_c h_{de}
      ],
\end{eqnarray}
\begin{equation}
N^{-1} D_a r^a = N_0^{-1} [\frac{2}{r} + \frac{2 h_{rr}}{r}
         + r_0^a\nabla_a( I+\frac{1}{2}h_{rr} ) ],
\end{equation}
\begin{equation}
r^a D_a N = - \frac{N_0}{2} r_0^a\nabla_a I
\end{equation}
and
\begin{equation}
\gamma^{-1/2}{\pi_a}^b r_b \sqrt{\sigma} =
   \frac{\sqrt{\sigma_0}}{2N_0}\frac{\partial h_{ar}}{\partial t}
      - \sqrt{\sigma_0} r_0^b \nabla_{(a} S_{b)}.
\label{defpiR}
\end{equation}

Now, consider the term proportional to the stationary $\delta S^a$ on
the left hand side of
Eq.~(\ref{masterST}).  It has no contribution from time dependent
terms:
all time dependent terms are multiplied by stationary terms and drop
out when integrated over a full period of~$\tau$.
With application of Eqs.~(\ref{defPi}, \ref{Kij},
\ref{riSi}) and use of the periodicity,
the terms proportional to $\delta S^a$ in
Eq.~(\ref{masterST}) contribute
\begin{eqnarray}
  \frac{N_0}{2\pi} \int_0^{2\pi} \oint_{r_B}
      2 \delta S^a \gamma^{-1/2} {\pi_a}^b r_b \sqrt{\sigma}
   \,d^2x\,d\tau
   = - \frac{N_0}{2\pi} \int_0^{2\pi} \oint_{r_B}
      \delta S^i (\nabla_i S_j + \nabla_j S_i) r_0^j
   \sqrt{\sigma_0} \,d^2x\,d\tau
\nonumber \\
   = \frac{N_0}{4\pi} \delta \left[
      \int_0^{2\pi}
      \oint_{r_B} r^{-1} S^i S_i \sqrt{\sigma_0} \,d^2x\,d\tau
   \right]
   - \frac{N_0}{2\pi} \int_0^{2\pi}
      \oint_{r_B} \delta S^i (\nabla_j S_i) r_0^j \sqrt{\sigma_0}
      \,d^2x\,d\tau .
\label{defSterm}
\end{eqnarray}
If there are no stationary sources at infinity, then
$S^i$ is of the form given in Eq.~(\ref{Sj}) and orthogonality
of the ${\cal Y}^{lm}_{K_l}$ implies that the
second term above is symmetric in $S^i$ and $\delta S^i$.
Thus the
right hand side of Eq.~(\ref{defSterm}) is
\begin{equation}
   \frac{N_0}{4\pi} \delta \left[
      \int_0^{2\pi} \oint_{r_B}
         ( S^i S_i /r - S^i r_0^j \nabla_j S_i )
      \sqrt{\sigma_0} \,d^2x\,d\tau
   \right] .
\label{deltaS}
\end{equation}

The other surface integral on the left hand side of
Eq.~(\ref{masterST}) is closely related to
\begin{eqnarray}
&& - \frac{1}{2\pi} \int_0^{2\pi} \oint_{r_B} \left[
      \delta (\sqrt{\sigma}\sigma^{ab})
         N (D_a r_b - \frac{1}{2} \sigma_{ab} D_c r^c)
      + \delta (N^2 \sqrt{\sigma}) N^{-1} D_a r^a
   + 2 \delta(\sqrt{\sigma}) r^a D_a N
   \right]\,d^2x\,d\tau
\nonumber \\ &&
\hspace{0.5in}  -\frac{N_0}{2\pi} \int_0^{2\pi} \oint_{r_B}
          \frac{1}{4} \delta ( h^{ij} r_0^k \nabla_k h_{ij})
      \sqrt{\sigma_0} \,d^2x\,d\tau.
\label{defST}
\end{eqnarray}
The part of expression~(\ref{defST}) depending upon the
stationary moments is
\begin{equation}
   \frac{N_0}{2\pi} \int^{2\pi}_0 \oint_{r_B}
      \delta I r_0^i \nabla_i I
   \sqrt{\sigma_0} \,d^2x\,d\tau .
\label{previous}
\end{equation}
And if there are no sources at infinity, then Eq.~(\ref{I}) and
the orthogonality of the ${\cal Y}^{lm}_{K_l}$ imply
that this is symmetric in $I$ and $\delta I$ and equal to
\begin{equation}
   \frac{N_0}{4\pi}
   \delta \left[
      \int^{2\pi}_0 \oint_{r_B}
         I r_0^i \nabla_i I
      \sqrt{\sigma_0} \,d^2x\,d\tau
   \right] .
\label{deltaI}
\end{equation}
Expressions~(\ref{deltaS}) and (\ref{deltaI}) are the
contributions from the stationary moments to the right hand side
of Eq.~(\ref{masterST}).

The radiative terms in expression~(\ref{defST})
follow with the substitutions from
Eqs.~(\ref{defsig}) to~(\ref{defpiR}) along with
Eqs.~(\ref{hii}), (\ref{Dihij}) and use of
Stoke's Theorem; the result is
\begin{eqnarray}
&&  \frac{N_0}{4\pi} \int_0^{2\pi} \oint_{r_B}
      (\frac{1}{2} \delta h^{ij} r_0^k \nabla_k h_{ij}
         - \frac{1}{2} h^{ij} r_0^k \nabla_k \delta h_{ij}
      ) \sqrt{\sigma_0} \,d^2x\,d\tau
\nonumber \\&&
   + \frac{N_0}{4\pi}
   \delta \left\{
      \int_0^{2\pi} \oint_{r_B}
      [- r_{0p} \nabla_i (h^{ij} \sigma_{0jq} h^{pq})
            + \frac{5}{2r} h_{rr} h_{rr}
            - \frac{1}{r} h_{ij} {h^i}_{k}\sigma_0^{jk}
      ] \sqrt{\sigma_0} \,d^2x\,d\tau
   \right\} .
\label{radST}
\end{eqnarray}
The focus remains on the first integral of
expression~(\ref{radST});
the second integral appears directly in Eq.~(\ref{masterST}).

The quantity $h_{ij}$ is decomposed into its $m,n$ components as
in Eq.~(\ref{hmn}), where $h^{mn}_{ij}$ is an exact solution to
the flat space wave equation~(\ref{hmnij}) with frequency
$\omega_{mn}$ and dependence upon $I^n_{lm}$, $\theta^{In}_{lm}$,
$S^n_{lm}$, $\theta^{Sn}_{lm}$ and also upon $\Omega$ and
$\omega_0$ through the dependence upon $\omega_{mn}$.  Thus
\begin{eqnarray}
   \delta h_{ij} & = & \sum_{lmn}
   \left[
      \frac{\delta h^{mn}_{ij}}{\delta I^{n}_{lm}}
            \delta I^{n}_{lm}
      + \frac{\delta h^{mn}_{ij}}{\delta \theta^{In}_{lm}}
            \delta \theta^{In}_{lm}
      + \frac{\delta h^{mn}_{ij}}{\delta S^{n}_{lm}}
            \delta S^{n}_{lm}
      + \frac{\delta h^{mn}_{ij}}{\delta \theta^{Sn}_{lm}}
            \delta \theta^{Sn}_{lm}
      + \frac{\delta h^{mn}_{ij}}{\delta \omega_{mn}}
            \delta \omega_{mn}
   \right] .
\label{dhij}
\end{eqnarray}

Now, the substitution of
Eq.~(\ref{dhij}) into the first integral of
expression~(\ref{radST}) and the use of
orthogonality properties of the
\( {\cal Y}^{lm}_{K_l} \) leads to a sum of terms each
involving a single choice of \(l,m,n\) and either $I^n_{lm}$
or $S^n_{lm}.$
The resulting terms that are proportional to
\( \delta I^{n}_{lm} \) and to \( \delta \theta^{In}_{lm} \) are
easy to evaluate because the coefficients of
\( \delta I^{n}_{lm} \) and \( \delta \theta^{In}_{lm} \) in
Eq.~(\ref{dhij}) are also solutions to Eq.~(\ref{hmnij}); and
the terms in the first integral of expression~(\ref{radST}),
which involve these,
form the Wronskian of two
solutions to Eq.~(\ref{hmnij}).  In fact, the term
involving \( \delta I^{n}_{lm} \) is just proportional to the
Wronskian of two dependent solutions of the same linear equation
and, hence, vanishes.  The surface integral of the Wronskian
involving \( \delta \theta^{In}_{lm} \) is independent of the
surface over which it is evaluated; and it is easiest to
evaluate in the wave zone.
The terms in the first integral of expression~(\ref{radST})
involving \( \delta S^{n}_{lm} \) and
\( \delta \theta^{Sn}_{lm} \) may be handled in a completely
analogous manner.
This results in
\begin{equation}
   N_0 \sum_{lmn} \left[
         \frac{8 (\omega_{mn})^{2l+1}(l+1)(l+2)}
                                          {(l!)^2 l (l-1)}
            I^n_{lm} I^{n\ast}_{lm} \delta \theta^{In}_{lm}
         + \frac{32 (\omega_{mn})^{2l+1} l (l+1)(l+2)}
                                          {[(l+1)!]^2 (l-1)}
            S^n_{lm} S^{n\ast}_{lm} \delta \theta^{Sn}_{lm}
      \right]
\label{deltatheta}
\end{equation}
for the \( \delta \)-phase contribution to the first integral of
expression~(\ref{radST}); this appears directly in
Eq.~(\ref{masterST}).

The term in the first integral of expression~(\ref{radST})
(after substitution from Eq.~(\ref{dhij})) which involves
\( \delta \omega_{mn} \) is rather more
complicated.  From Eq.~(\ref{hmnij})
it is clear that the product
\( r h^{mn}_{ij} \) may be written so that $r$ and
$\omega_{mn}$ only appear in the combination $r \omega_{mn}$,
except for a possible overall amplitude dependence upon
$\omega_{mn}$.
Thus
\begin{equation}
   \delta \omega_{mn}
      \frac{\delta h^{mn}_{ij}} {\delta \omega_{mn}}
      = \delta \omega_{mn}
          \frac{\delta (r h^{mn}_{ij})}{\delta(r \omega_{mn})},
\end{equation}
and \( \delta (r h^{mn}_{ij}) / \delta(r \omega_{mn}) \)
is closely related to just the $r$ derivative of
\(r h^{mn}_{ij} \), so that
\begin{equation}
   \delta \omega_{mn}
      \frac{\delta h^{mn}_{ij}} {\delta \omega_{mn}}
   =  \frac{\delta \omega_{mn}}{\omega_{mn}}
          \frac{\partial}{\partial r} (r h^{mn}_{ij})
   =  \frac{\delta \omega_{mn}}{\omega_{mn}}
          r_0^k \nabla_k (r h^{mn}_{ij}).
\end{equation}
Now, the \( \delta \omega_{mn} \) contribution to the first
integral of expression~(\ref{radST}) is
(let \( h_{ij} \rightarrow h^{{mn}\ast}_{ij} \) and
   \( \delta h_{ij} \rightarrow
      \delta \omega_{mn}
      r_0^k \nabla_k (r h^{mn}_{ij}) / \omega_{mn}
   \),
sum over $m$ and $n$ and do the integral over $\tau$)
\begin{equation}
   \frac{N_0}{4} \sum_{mn} \oint_{r_B}\
      \frac{\delta \omega_{mn}}{\omega_{mn}}
      \{ r_0^p \nabla_p (r h^{mn}_{ij})
               r_0^q \nabla_q h^{{mn}\ast}_{ij}
         - h^{{mn}\ast}_{ij} r_0^p \nabla_p
               [r_0^q \nabla_q (r h^{mn}_{ij})]
      \} \sqrt{\sigma_0}\,d^2x .
\end{equation}
The term in braces simplifies, and a
\( \delta \)-variation by parts and average over $\tau$ result
in
\begin{eqnarray} &&
   \frac{N_0}{8\pi}
   \delta \left\{ \int_0^{2\pi} \oint_{r_B}
         [ r r_0^p \nabla_p (h_{ij}) r_0^q \nabla_q
                                       h^{ij}
            - h_{ij} r_0^p \nabla_p
                  (r r_0^q \nabla_q h^{ij})
         ] \sqrt{\sigma_0} \,d^2x\,d\tau
      \right\}
\nonumber\\&&
   - \frac{ N_0}{8\pi} \sum_{{mn}}
            (n \omega_0 - \frac{m \Omega}{N_0})
      \delta \left\{
         \int_0^{2\pi} \oint_{r_B} \omega_{mn}^{-1}
         [ r r_0^p \nabla_p (h^{mn}_{ij}) r_0^q \nabla_q
                                       h^{{mn}\ast}_{ij}
      \right.
\nonumber\\&&
      \left.
            - h^{{mn}\ast}_{ij} r_0^p \nabla_p
                  (r r_0^q \nabla_q h^{mn}_{ij})
         ] \sqrt{\sigma_0} \,d^2x\,d\tau
      \right\};
\label{deltaomega}
\end{eqnarray}
this appears directly in Eq.~(\ref{masterST}).


\begin{thebibliography}{1}

\bibitem{ThorneRMP}
K.~S. Thorne, Reviews of Modern Physics {\bf 52},  299  (1980).

\bibitem{MTBH}
S. Chandrasekhar, {\em The Mathematical Theory of Black Holes}
   (Oxford University Press, Oxford, 1983).

\bibitem{ThorneII}
R. Price and K.~S. Thorne, Astrophysical Journal {\bf 155},
   163  (1969).

\bibitem{gibbons84}
   G.~W. Gibbons and J.~M. Stewart, in {\em Classical General
   Relativity}, edited by W.~B. Bonner, J.~N. Islam and
   M.~A.~H.  MacCallum (Cambridge University Press, Cambridge,
   1984), pp.\ 77--94.

\bibitem{paperI}
J.~K. Blackburn and S. Detweiler, Physical Review D {\bf 46},
   2318  (1992).

\bibitem{ThorneIII}
   K.~S. Thorne, Astrophysical Journal {\bf 158},  1  (1969).

\bibitem{YorkQLE}
J.~D. Brown and J.~W. York, Physical Review D {\bf 47},
   1407  (1993).

\bibitem{YorkIV}
J.~W. York,  in {\em Sources of Gravitational Radiation},
   edited by L. Smarr
   (Cambridge University Press, Cambridge, 1979), pp.\ 83--126.

\bibitem{ADM}
R. Arnowitt, S. Deser, and C.~W. Misner,  in {\em Gravitation:
   an Introduction to Current Research}, edited by L. Witten
   (John Wiley \& Sons, New York,
   1962), Chap.~7, pp.\ 227--265.

\bibitem{hayward93}
G. Hayward, Physical Review D {\bf 47}, 3275 (1993).

\bibitem{ReggeandT}
T. Regge and C. Teitelboim, Annals of Physics {\bf 88},
   286  (1974).
\end{thebibliography}
\end{document}